**Copula-Based Reconstruction and Clustering of Coccidioides Minimum Inhibitory Concentration Profiles**

MIRANDA WASCO[1], PAROMITA BANERJEE[2*]

1 Department of Biology, John Carroll University, Cleveland, Ohio, USA

2 Department of Mathematics, Computer Science, and Data Science, John Carroll University, Cleveland, Ohio, USA

*Corresponding author: pbanerjee@jcu.edu

Paromita Banerjee https://orcid.org/0000-0002-6218-615X

**Abstract**

Coccidioidomycosis is a fungal lung infection endemic to parts of the Pacific Northwest and southwestern United States, Mexico, Central America, and South America. A 2017 study by Thompson et al. reported that many analyzed *Coccidioides* isolates had elevated minimum inhibitory concentration values for fluconazole but comparatively low minimum inhibitory concentration values for other triazole drugs. We constructed 172 synthetic joint minimum inhibitory concentration profiles from the published drug-specific marginal frequency tables. Values sampled from each marginal distribution were paired across drugs using a Gaussian copula, and potential cross-drug patterns were examined using k-means clustering, hierarchical clustering, and Gaussian mixture models. On the selected reconstruction, a forced four-cluster partition consistently identified an upper-caspofungin minimum-inhibitory-concentration tail. Across the evaluated dependence scenarios, the gap statistic favored a single cluster in 96% to 100% of nested reconstructions, providing no evidence that the reconstructed multivariate data supported a broader multicluster structure. A separate simulation study examined recovery of an upper-severity-score category defined from prespecified quantiles of a composite minimum-inhibitory-concentration score. The study included 4,000 replications across 20 prevalence and measurement-noise conditions. The multiclass adjusted Rand index ranged from moderate to high depending on the method and prevalence, whereas binary partition agreement for the upper-severity-score category approached zero at 1% prevalence even though recall remained near 1.0. Clustering conclusions therefore depended on the assumed cross-drug dependence structure and on the realized number of observations in the target category.

## 1. Introduction

Coccidioidomycosis, also known as Valley Fever, is a lung infection caused by the soil-dwelling fungi Coccidioides immitis and Coccidioides posadasii. These species are endemic to parts of the Pacific Northwest, the southwestern United States, Mexico, Central America, and South America.[1] Although approximately 60% of infections are asymptomatic, about 5-10% develop into severe or long-term pulmonary disease, and approximately 1% result in disseminated infection.[1,2] The risk of severe disease is higher among adults aged 60 years or older, immunocompromised individuals, pregnant women, people with diabetes, and members of African American and Filipino populations.[2] A 2025 modeling study estimated 273,000 incident symptomatic cases in the United States in 2019 (95% credible interval [CrI], 206,000-360,000), corresponding to a national incidence of approximately 58 to 117 cases per 100,000 people, depending on regional endemicity.[3] This estimate was 10-18 times the 20,003 cases reported in 2019, reflecting underreporting and missed diagnoses.[3,4] Incidence has also risen in association with travel to endemic areas, recreational and occupational exposure, increasing comorbidity, climate change, and improved detection.[4]

The 2016 Infectious Diseases Society of America (IDSA) clinical practice guideline states that most immunocompetent individuals with coccidioidal pneumonia overcome the infection without antifungal treatment.[5] When treatment is indicated, the choice of therapy and patient outcomes depend on pharmacokinetic and pharmacodynamic properties, host characteristics, disease severity, drug administration, pathogen susceptibility, adverse effects, and dosage.[6] Fluconazole, an oral azole, is commonly prescribed for uncomplicated coccidioidomycosis because it is well absorbed, has relatively few drug-drug interactions, and is comparatively inexpensive.[5] A 2025 study reported that 87.4% of patients treated between 2010 and 2024 initially received fluconazole monotherapy, confirming its common use as initial therapy.[7]

However, Thompson et al. reported elevated fluconazole MIC values: 37.3% of the analyzed isolates had values of at least 16 μg/mL, and 3.8% had values of at least 64 μg/mL.[8] Most isolates with very high fluconazole MIC values (≥64 μg/mL) had comparatively low MIC values for the other triazoles.[8] The clinical significance of these findings remains difficult to interpret because validated Coccidioides-specific breakpoints are unavailable.[9]

Clustering of MIC profiles

Because the published frequency tables report drug-specific marginal distributions rather than joint isolate-level measurements, cross-drug MIC patterns cannot be observed directly, motivating two related questions addressed in this study: whether synthetic joint MIC profiles reconstructed from these marginal distributions support stable multivariate clusters, and how reliably clustering methods recover a prespecified high-MIC category when that category is rare. Both questions were evaluated across reconstruction assumptions, target prevalence, and realized target counts, factors expected to influence whether a rare high-MIC profile is recovered at all. Because successful recovery of a predefined category does not by itself indicate natural population structure, cluster stability and category recovery were assessed as distinct outcomes throughout. Any inferred cross-drug pattern therefore remains conditional on the assumed dependence structure, and confirmation would require joint MIC measurements from individual isolates.

## 2. Materials and Methods

This study analyzed synthetic joint *Coccidioides* minimum inhibitory concentration (MIC) profiles constructed from drug-specific marginal frequency distributions reported by Thompson et al. in "Large-Scale Evaluation of *In Vitro* Amphotericin B, Triazole, and Echinocandin Activity against Coccidioides Species from U.S. Institutions." The analysis included the five drugs with the most complete published data: fluconazole (n = 581), itraconazole (n = 486), posaconazole (n = 377), voriconazole (n = 499), and caspofungin (n = 172).[8] For each of the four larger marginals, 172 values were sampled without replacement from the full published frequency distribution (Table 2 in Thompson et al.8); caspofungin required no resampling because its published sample size was 172. Open-ended top bins (fluconazole ≥64 µg/mL; itraconazole and posaconazole >16 µg/mL) were assigned their boundary concentrations. Because the marginal frequencies do not identify which drug values occurred in the same isolate, the reconstruction sensitivity analysis evaluated uncertainty in the pairing of values across drugs.

Throughout the descriptive clustering analysis, k = 4 was used as a common partition size for comparison across k-means, hierarchical clustering, and Gaussian mixture modeling (GMM). The gap statistic was used separately to evaluate the number of groups supported by the reconstructed data.

K-means clustering grouped profiles according to similarity in MIC values. This unsupervised method partitions observations into a prespecified number of clusters, k, by minimizing within-cluster variance. We used the Hartigan-Wong algorithm implemented in R, with each cluster represented by the mean, or centroid, of its member profiles.[10,11] Principal component analysis (PCA) displayed the k-means clusters in two dimensions using the first two principal components. A heatmap created with the pheatmap R package summarized cluster-level MIC patterns across the five drugs.[19] Overall and cluster-specific average silhouette widths measured cluster separation.[10,12] Bootstrap stability was summarized by the maximum Jaccard coefficient between each original cluster and its best-matching cluster across bootstrap resamples.[13,30]

The adjusted Rand index (ARI) and confusion matrices quantified agreement between clustering methods applied to the same reconstruction. The ARI was not treated as external validation because observed joint profiles and verified cluster labels were unavailable. Gaussian mixture models assigned each profile a probability of membership in each of G components. A separate resampling experiment evaluated recovery of a prespecified upper-severity-score category across varying prevalence and dilution-noise levels using the median F1 score, precision, recall, false-positive rate, and ARI. The F1 score is the harmonic mean of precision and recall.

### 2.1. Study Design and Outcomes

The analyses had two connected aims. First, the reconstruction analysis assessed the number and stability of multivariate groups and evaluated whether a forced partition could isolate an upper-caspofungin MIC pattern. Second, the rare-category experiment examined how target prevalence and dilution noise affected recovery of a prespecified composite upper-severity-score category. These targets were defined independently of the fitted cluster assignments and were evaluated as descriptive categories, not as validated clinical resistance classes.

The reconstruction analysis used two reference targets. A cluster-level upper-caspofungin MIC pattern was identified from cluster medians, and isolate-level recovery was evaluated against a raw caspofungin MIC threshold of ≥2 µg/mL. The rare-category experiment used the composite upper-severity-score category defined across all five drugs. None of these descriptive targets represents validated clinical antifungal resistance because *Coccidioides*-specific breakpoints are unavailable.[9]

*Table 1. Overview of the reconstruction and replication designs.*

| Design | Reconstructions / replications | What varies | Reported in |
|---|---|---|---|
| Nested reconstruction analysis | 800 (25 outer x 8 inner x 4 dependence scenarios) | Both which 172 values are sampled per drug (outer) and how they are paired across drugs (inner) | Section 3.2, summary statistics |
| Fixed-marginal repeated-pairing analysis | 1,000 (200 pairings x 5 dependence scenarios) | Only the pairing across drugs; the 172 marginal values are fixed to one sample | Section 3.2, Table 3 |
| Sample-size sensitivity | 450 (3 sizes x 3 dependence conditions x 50 replications) | Reconstruction size (n = 172, 300, 500) and dependence condition | Section 3.2 |
| Open-bin encoding sensitivity | 400 (2 encodings x 4 dependence scenarios x 50 replications) | Open-bin encoding convention and dependence scenario | Section 3.2 |
| Rare-category recovery experiment | 4,000 (5 prevalence levels x 4 noise levels x 200 replications) | Target prevalence and dilution-noise level, with profiles drawn from the selected reconstruction | Section 3.3, Table 4 |

## 2.2. Reconstruction Sensitivity Analysis

A reconstruction sensitivity analysis examined uncertainty in the unknown pairing of drug values within each synthetic profile. Each drug's published marginal MIC distribution was held fixed while joint dependence was varied with a Gaussian copula,[14] producing datasets with independent, weak ($\rho = 0.3$), moderate ($\rho = 0.6$), and strong ($\rho = 0.85$) positive dependence. Near-perfect rank alignment was included as an extreme sensitivity condition. At $k = 4$, each dataset was clustered using k-means, Ward.D2 hierarchical clustering, and GMM. A cluster-level upper-caspofungin MIC pattern was recorded when one cluster's median caspofungin MIC exceeded the next-highest cluster median by at least fivefold and included fewer than half of the profiles. Two hundred datasets were generated per dependence level for k-means, hierarchical clustering, and GMM. For the descriptive results in Section 3.1, one reconstruction was selected from the 800 nested reconstructions. Selection minimized the Euclidean distance from the joint median of the standardized median silhouette width and standardized median target-recovery F1 score, with equal weights. Ties were resolved using the lowest outer-sample index, followed by the lowest inner-pairing index. Isolate-level recovery was measured against the raw caspofungin MIC threshold of ≥2 µg/mL defined in Section 2.1. This F1 score was used to select the reconstruction and in the analyses reported in Section 3.2. The separate fivefold-median criterion indicated whether a cluster-level upper-caspofungin MIC pattern was present. K-means clustering used 25 random starts (nstart = 25) throughout the study. A caspofungin-ablation analysis examined category recovery when caspofungin was omitted, and the gap statistic[15] assessed the number of clusters supported by the reconstructed data.

## 2.3. Rare-Category Recovery Experiment

Each of the 4,000 replications contained 500 complete synthetic MIC profiles sampled with replacement from the selected 172-row reconstruction. The four severity-score categories were quartiles of that reconstruction, so each contained approximately 25% of its 172 profiles. Target prevalence was imposed during resampling at 1%, 2%, 5%, 10%, or 15%. Profiles were drawn with unequal probability from the upper-severity-score quartile and the other three quartiles combined so that the expected share of the upper category in each 500-profile replication matched the specified prevalence. The design crossed five target-prevalence levels with four dilution-noise levels, with 200 replications in each of the 20 conditions. Cross-drug dependence, separation between profiles, and any extreme observations were inherited from the selected reconstruction.

Clustering of MIC profiles

Before clustering, each sampled profile was assigned to one of four severity-score categories using deterministic quartile cutoffs for the mean standardized log2 MIC across the five drugs. The highest category was the prespecified upper-severity-score target. The experiment used k = 4 to evaluate recovery of all four score categories, with sensitivity analyses considering k = 2 through 6. Predicted clusters were ordered by mean standardized fluconazole MIC and matched by rank to the ordered severity-score categories.

For a replication assigned noise level $\varepsilon \in \{0, 0.05, 0.10, 0.20\}$, each drug's log2 MIC value in each resampled profile was independently shifted by -1, 0, or +1, representing one twofold-dilution step down, no change, or one step up. The corresponding probabilities were $\varepsilon/2$, $1-\varepsilon$, and $\varepsilon/2$. Noise was generated independently across drugs and profiles and was distinct from the cross-drug dependence imposed during reconstruction. Shifted values were restricted to each drug's observed [minimum - 2, maximum + 2] log2 range before conversion back to raw MIC. When $\varepsilon = 0$, no dilution noise was added.

2.4. Gap Statistic

For each candidate k from 1 through 6, observed within-cluster dispersion was compared with B = 50 Monte Carlo reference datasets drawn uniformly from the bounding box of the observed data in the same scaled log2 feature space used for clustering. The analysis used the clusGap function in the R package cluster. The selected k was the smallest value satisfying the one-standard-error rule of Tibshirani et al.[15] (method = "Tibs2001SEmax" in clusGap's maxSE function), rather than the value that simply maximized the raw gap statistic.

2.5. Gaussian Mixture Modeling

All Gaussian mixture models were fitted with the R package mclust. Fixed-component analyses used model EII, which assumes diagonal, equal-volume, spherical covariance across components. The number of components, G, was specified for the descriptive clustering, reconstruction sensitivity, and rare-category recovery analyses. For the Bayesian information criterion (BIC) comparison reported in Results, mclust evaluated G = 1 through 9 across the available covariance models and selected the model with the maximum BIC. Initialization used the package's default hierarchical model-based agglomeration values. The expectation-maximization (EM) algorithm used the default convergence tolerance (relative log-likelihood change < 1 × 10^-5) and a maximum of 1,000 iterations.

2.6. Software

All analyses were conducted in R[16] using the packages cluster[17], factoextra[18], pheatmap[19], and mclust[20] for clustering, model fitting, and cluster visualization. Data manipulation and figure generation used ggplot2[21], dplyr[22], purrr[23], tidyr[24], tibble[25], and reshape[26], collectively part of the tidyverse[27]. Model comparison utilities were drawn from caret[28], plotting palettes from gcookbook[29], and bootstrap stability estimation from bootcluster[30].

## 3. Results

### 3.1. Descriptive Clustering of a Selected Reconstruction

A forced k = 4 partition was applied to the selected reconstruction. The between-cluster sum of squares was 47.3% of the total sum of squares, and raw MIC medians were calculated for each drug within each cluster.

*Table 2. Cluster medians of the raw data from a forced k = 4 partition on the selected reconstruction (µg/mL)*

| | Cluster Size | FLC | ITR | POS | VOR | CAS |
|---|---|---|---|---|---|---|
| Cluster 1 | 23 | 0.5 | 0.25 | 0.12 | 0.06 | 0.12 |
| Cluster 2 | 49 | 8 | 0.12 | 0.06 | 0.06 | 0.12 |
| Cluster 3 | 73 | 16 | 0.5 | 0.25 | 0.12 | 0.12 |
| Cluster 4 | 27 | 16 | 0.5 | 0.25 | 0.12 | 8 |

*FLC, fluconazole; ITR, itraconazole; POS, posaconazole; VOR, voriconazole; CAS, caspofungin.*

Cluster 1 (n = 23) had the lowest overall MIC values in the forced partition. Clusters 2 and 3 (n = 49 and 73) had median fluconazole MIC values of 8 and 16 µg/mL, respectively, with low-to-moderate values for the other drugs; both had a median caspofungin MIC of 0.12 µg/mL. Clusters 3 and 4 had identical medians for every drug except caspofungin. Cluster 4 (n = 27) had a median caspofungin MIC of 8 µg/mL, approximately 67-fold higher than the other three clusters.

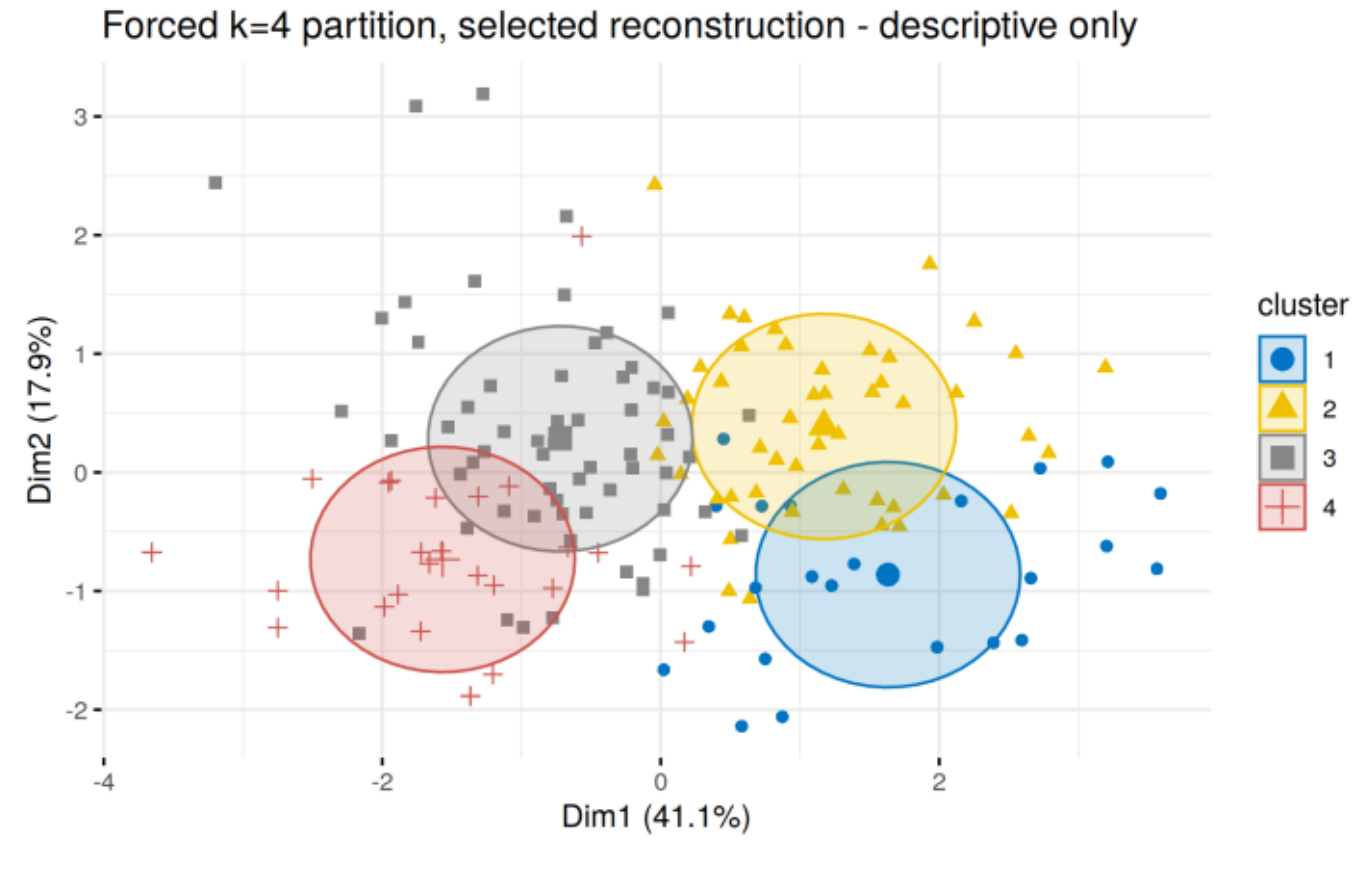

*Fig. 1. Principal component analysis (PCA) plot for the forced k = 4 partition on the selected reconstruction. PC1 explains 41.1% of variance; PC2 explains 17.9%.*

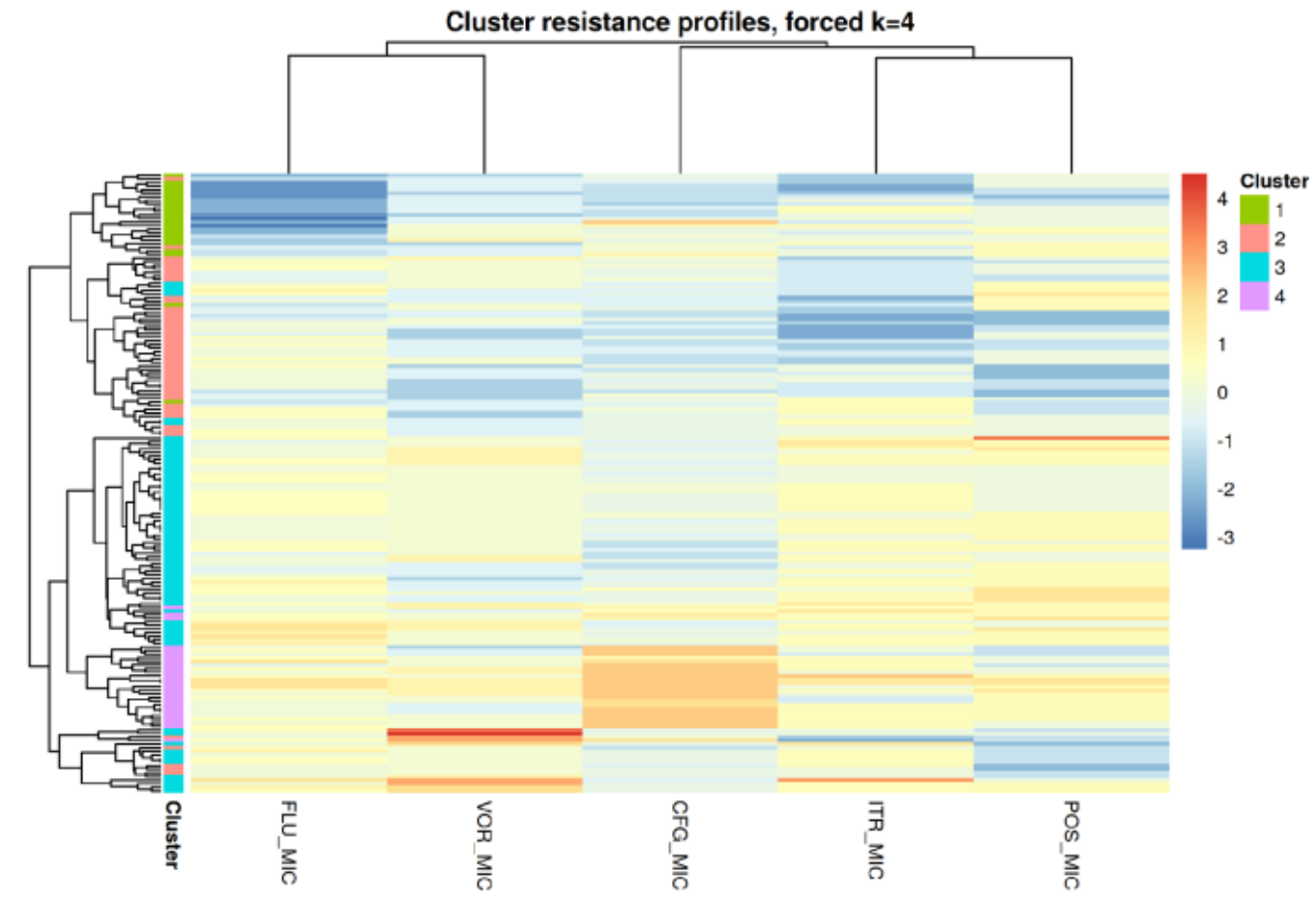

*Fig. 2. Heatmap of cluster-level MIC patterns for the forced k = 4 partition on the selected reconstruction.*

The four groups overlapped substantially in the selected reconstruction. PC1 explained 41.1% of the variance, compared with 87.9% under the extreme near-perfect-alignment condition.

The heatmap shows a distinct caspofungin stripe for Cluster 4 but otherwise weak differentiation among clusters across the other four drugs, consistent with Table 2's near-identical medians for Clusters 3 and 4 outside of caspofungin.

Cluster stability was assessed with a nonparametric bootstrap (B = 500) using the maximum Jaccard coefficient between each original cluster and its best-matching counterpart in each resample. The overall mean Jaccard coefficient was 0.498 (cluster means: 0.595, 0.450, 0.512, and 0.436; interquartile ranges: [0.55, 0.67], [0.38, 0.53], [0.46, 0.58], and [0.26, 0.61]). Dissolution frequencies, defined as the proportion of resamples with a best-match Jaccard coefficient below 0.5, were 15.2%, 66.6%, 39.8%, and 55.8% for Clusters 1 through 4, respectively. The overall average silhouette width was 0.25, with cluster-specific widths of 0.21, 0.20, 0.28, and 0.29, compared with 0.53 under near-perfect alignment.

The bootstrap quantified row-resampling stability for the selected reconstruction and was not repeated across the 800-reconstruction sensitivity analysis.

A Gaussian mixture model was fit to the selected reconstruction with G = 4 and model EII. Agreement between k-means and GMM at G = 4 was weak (ARI = 0.316). When G was selected by BIC, four components were selected for the selected reconstruction and two under near-perfect alignment, indicating sensitivity to the assumed dependence structure.

Clustering of MIC profiles

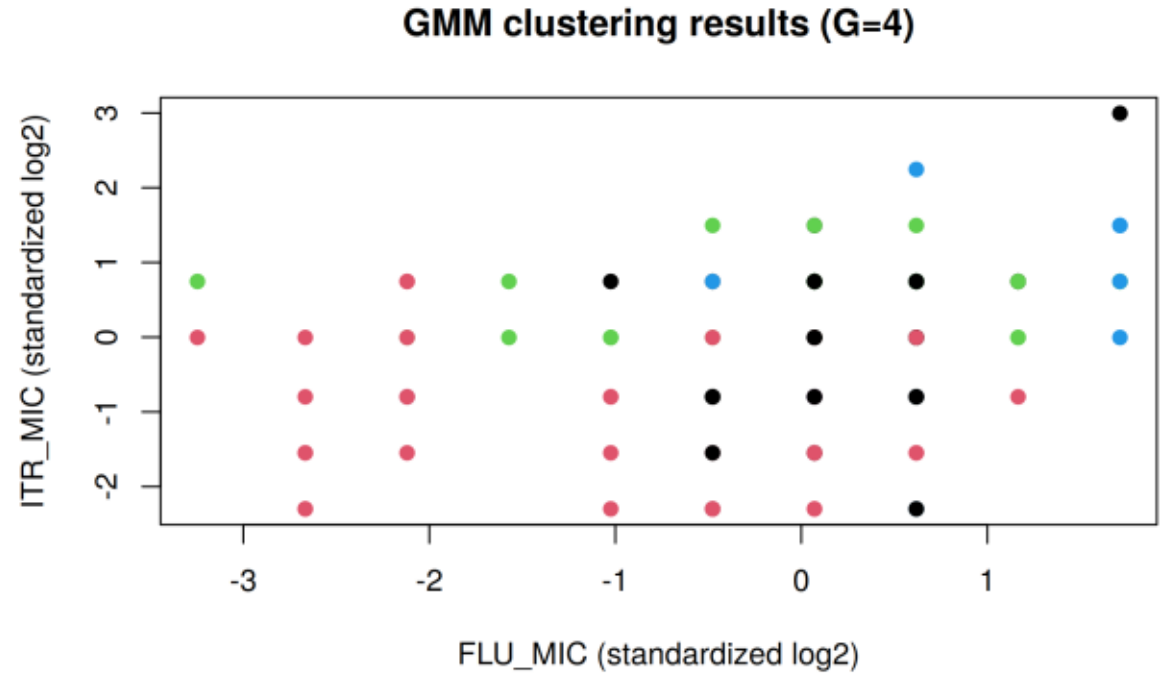

*Fig. 3. Gaussian mixture model (GMM) clustering results for fluconazole and itraconazole minimum inhibitory concentrations at G = 4 on the selected reconstruction.*

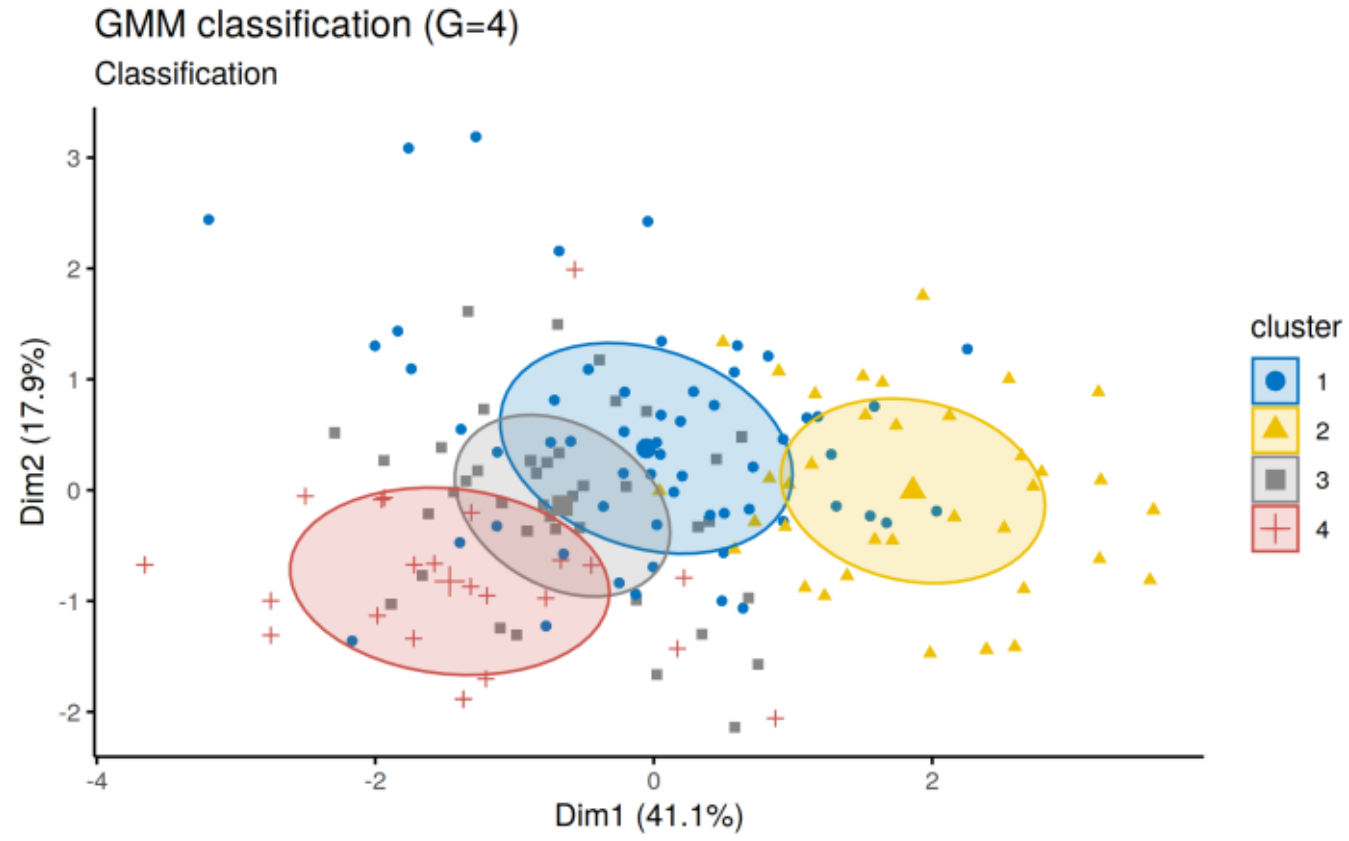

*Fig. 4. Gaussian mixture model (GMM) classification cluster plot at G = 4 on the selected reconstruction.*

3.2. Reconstruction Uncertainty and Sensitivity Analyses

A nested design evaluated marginal-sampling and pairing uncertainty. 25independent outer marginal samples were each subjected to 8 inner pairing reconstructions under 4dependence scenarios: independent, weak ($\rho$ = 0.3), moderate ($\rho$ = 0.6), and strong ($\rho$ = 0.85). This produced 800 nested reconstructions.

Across all 800 nested reconstructions, the gap statistic selected k = 1 in 96-100% of reconstructions within every dependence scenario: 200 of 200 under independence and strong dependence, 193 of 200 under moderate dependence, and 192 of 200 under weak dependence. The remaining reconstructions selected k = 2-4. In the moderate-dependence k-means condition, the within-outer-sample variance in target-recovery F1 attributable to redrawing the pairing was 0.0025, whereas the between-outer-sample variance attributable to redrawing the 172 marginal values was 0.0003. Pairing uncertainty was therefore about eight times larger than marginal-sampling uncertainty in this condition. Median F1 for recovery of the upper-caspofungin MIC target remained between 0.90 and 0.96 across methods and dependence scenarios.

*Table 3. Percentage of reconstructions in which a distinct upper-caspofungin MIC tail was detected at k = 4, by clustering method and dependence scenario*

| Method | Independent | Weak (ρ=0.3) | Moderate (ρ=0.6) | Strong (ρ=0.85) | Near-perfect alignment |
|---|---|---|---|---|---|
| k-means | 100.0% | 100.0% | 100.0% | 100.0% | 100.0% |
| Hierarchical (Ward.D2) | 100.0% | 100.0% | 100.0% | 100.0% | 100.0% |
| GMM | 99.0% | 100.0% | 99.5% | 99.5% | 100.0% |

Table 3 reports a secondary analysis of upper-caspofungin MIC tail recovery at k = 4. One marginal sample was held fixed while 200 pairings were generated within each of 5 dependence scenarios, from independence to near-perfect rank alignment, for a total of 1,000 pairing reconstructions. In the 800-reconstruction nested analysis, each scenario likewise included 200 reconstructions, with both the 172-value marginal samples and cross-drug pairings regenerated. The upper-caspofungin MIC tail was recovered in essentially every reconstruction by all three methods. Against the isolate-level threshold of raw caspofungin MIC ≥2 µg/mL, median precision was 0.81 to 0.93, median recall was 0.92 to 1.0, median F1 was 0.89 to 0.94, and the median false-positive rate was 0.01 to 0.04. Near 100% detection of the cluster-level pattern did not imply perfect isolate-level classification because some profiles near the threshold were misclassified. Full distributions and 95% simulation intervals are provided in the supplementary results. Performance was similar across the evaluated dependence conditions.

The gap statistic generally selected k = 1, while a forced partition recovered the caspofungin-defined target when caspofungin was included. In the ablation analysis, clustering with the other four drugs produced median

F1 values of 0.17 to 0.67, compared with 0.92 to 0.94 when caspofungin was included. Median F1 was 0.17 to 0.27 under independent or weak pairing and 0.51 to 0.67 under moderate or strong assumed dependence. Target recovery was therefore driven primarily by caspofungin itself rather than by a stable multivariate pattern across the remaining drugs.

A sample-size sensitivity analysis examined whether the findings depended on the reconstruction size of n = 172, corresponding to the published caspofungin sample. 450reconstructions were generated ($n \in \{172, 300, 500\}$ × independent, moderate ($\rho = 0.6$), and a heterogeneous dependence condition × 50 replications each), drawing with replacement from the published marginal distributions at each sample size. The heterogeneous condition used $\rho = 0.6$ among the four triazoles and $\rho = 0.2$ between each triazole and caspofungin. The gap statistic selected k = 1 in 98% or 100% of reconstructions in every sample size and dependence condition cell (99.3% overall, 95% binomial confidence interval 98.1-99.9%). Median target-recovery F1 for the upper-caspofungin MIC tail was similar across sizes (0.91 at n = 172, 0.91 at n = 300, 0.92 at n = 500; full range 0.78-1.00 across all 450 reconstructions).

An open-bin encoding sensitivity analysis compared boundary encoding of the top MIC bins (fluconazole ≥64 µg/mL; itraconazole and posaconazole >16 µg/mL) with encoding one twofold-dilution step above each boundary (128, 32, and 32 µg/mL, respectively). The analysis included four dependence scenarios and 50 reconstructions per encoding-by-scenario cell (400 reconstructions total, n = 172). The gap statistic selected k = 1 in 94% to 100% of reconstructions under both encodings (98.5% overall, 95% binomial confidence interval 96.8% to 99.4%). Median target-recovery F1 was 0.92 under both encodings, with a full range of 0.20 to 1.00. A separate comparison of B = 50 and B = 200 reference samples on 20 held-out reconstructions selected the same k in every case. Results were similar in the sample-size sensitivity analysis and in the separate open-bin encoding sensitivity analysis.

### 3.3. Rare-Category Recovery Experiment

The rare-category analysis comprised 4,000 replications, with 200 replications for each of the 20 prevalence-by-noise conditions. Each replication contained 500 profiles and was analyzed using k-means, Ward.D2 hierarchical clustering, and GMM at k = 4. Performance was assessed against the prespecified severity-score labels using the F1 score, precision, recall, false-positive rate, and ARI. All three methods completed every replication without a nonconvergent fit or an empty predicted cluster.

*Table 4. Median F1 score, precision, recall, false-positive rate, and adjusted Rand index at 0% dilution noise across five prevalence levels, by clustering method*

| Method | Prev. | Noise | Med. F1 | Med. Prec. | Med. Recall | Med. FPR | Med. ARI |
|---|---|---|---|---|---|---|---|
| GMM | 1% | 0% | 0.105 | 0.056 | 1.000 | 0.136 | 0.407 |
| GMM | 2% | 0% | 0.957 | 1.000 | 1.000 | 0.000 | 0.416 |
| GMM | 5% | 0% | 0.966 | 1.000 | 0.933 | 0.000 | 0.424 |
| GMM | 10% | 0% | 0.968 | 1.000 | 0.938 | 0.000 | 0.454 |
| GMM | 15% | 0% | 0.971 | 1.000 | 0.943 | 0.000 | 0.487 |
| Hierarchical | 1% | 0% | 0.056 | 0.029 | 1.000 | 0.333 | 0.868 |
| Hierarchical | 2% | 0% | 0.111 | 0.059 | 1.000 | 0.322 | 0.852 |
| Hierarchical | 5% | 0% | 0.979 | 1.000 | 0.961 | 0.000 | 0.488 |
| Hierarchical | 10% | 0% | 0.989 | 1.000 | 0.978 | 0.000 | 0.521 |
| Hierarchical | 15% | 0% | 0.986 | 1.000 | 0.973 | 0.000 | 0.551 |
| k-means | 1% | 0% | 0.062 | 0.032 | 1.000 | 0.310 | 0.702 |
| k-means | 2% | 0% | 0.960 | 1.000 | 1.000 | 0.000 | 0.421 |
| k-means | 5% | 0% | 0.970 | 1.000 | 0.941 | 0.000 | 0.426 |
| k-means | 10% | 0% | 0.972 | 1.000 | 0.945 | 0.000 | 0.455 |
| k-means | 15% | 0% | 0.971 | 1.000 | 0.944 | 0.000 | 0.487 |

*Prev., prevalence of the upper-severity-score category; Med. F1, median F1 score; Med. Prec., median precision; Med. FPR, median false-positive rate; Med. ARI, median adjusted Rand index.*

Complete results across all prevalence and noise conditions, including 95% simulation intervals based on the 2.5th and 97.5th percentiles of replicate-level performance, are reported in Table S1 of the supplementary materials.

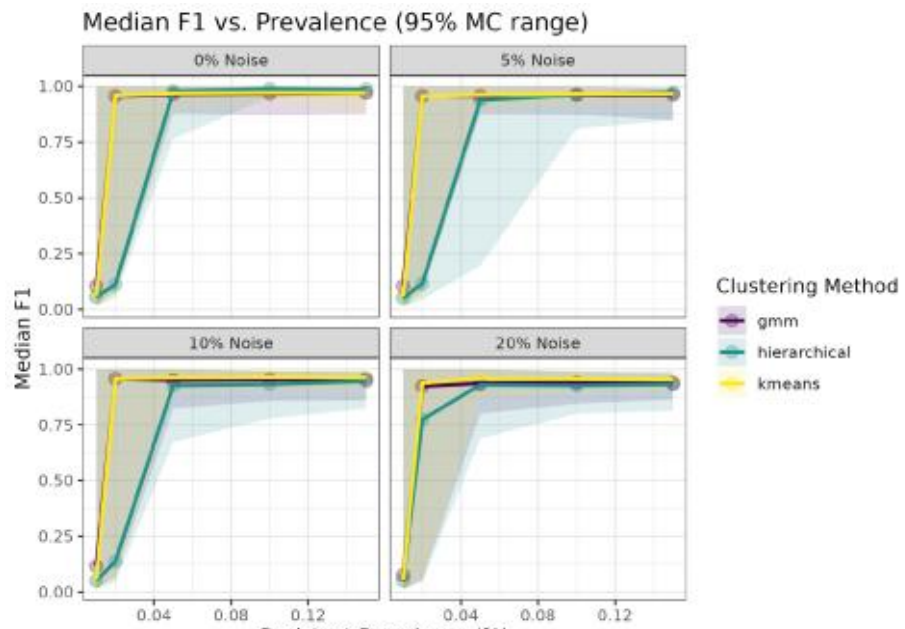

*Fig. 5. Median F1 score by prevalence of the resampled upper-severity-score category, faceted by noise level, with 95% simulation interval (4,000 replications).*

*Fig. 6. Median recall by prevalence, faceted by noise level.*

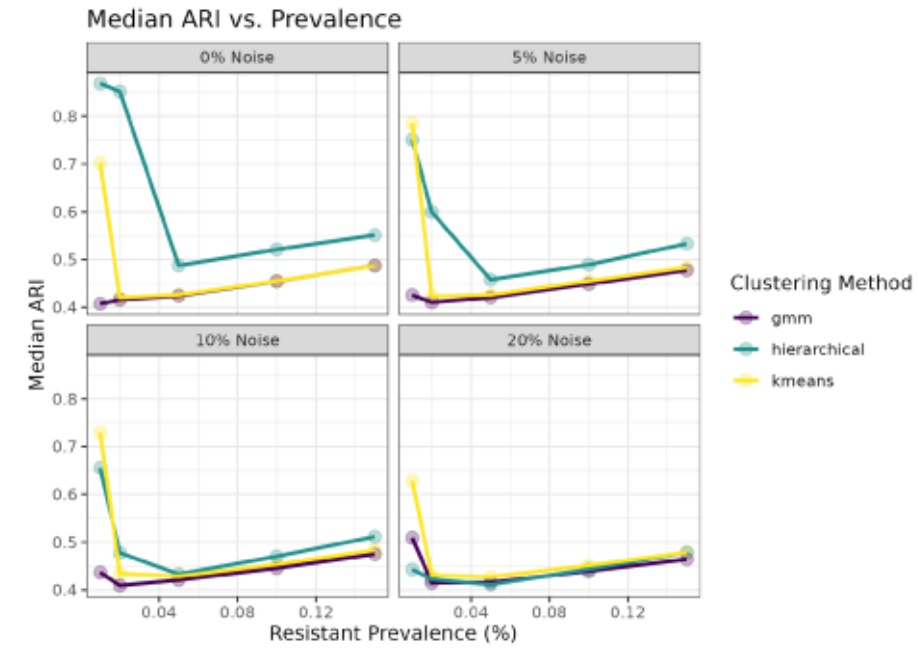

*Fig. 7. Median adjusted Rand index by prevalence, faceted by noise level.*

Recall remained near 1.0 for all methods across prevalence and noise conditions, and the multiclass ARI remained moderate to high. In contrast, precision and the F1 score declined sharply at low prevalence: median F1 was as low as 0.05 at 1% prevalence but ranged from 0.93 to 0.99 at 15% prevalence. Thus, the predicted upper-severity-score clusters captured most target profiles but included many false positives when the target was rare. Hierarchical clustering showed an unusual pattern at 2% prevalence: median F1 increased from approximately 0.11 to 0.14 at 0% to 10% noise to 0.77 at 20% noise. The distribution was bimodal at every noise level, and the proportion of successful replications increased from 32% to 39.5%, 47.5%, and 53.5% as noise increased. This pattern should be interpreted cautiously because it may reflect changes in partition geometry under the simulated perturbations rather than a general benefit of measurement noise.

Performance was more closely related to the realized number of target profiles than to nominal prevalence alone. In a 500-profile replication, the realized target count varied around 5 at 1% nominal prevalence and 10 at 2% nominal prevalence. At 0% noise, pooling the 200 replications at 1% prevalence with the 200 at 2% prevalence (400 replications total), the median k-means F1 score ranged from 0.01 to 0.07 when six or fewer target profiles were realized and from 0.92 to 1.0 when seven or more were realized. GMM showed a similar transition near six target profiles, whereas hierarchical clustering improved more gradually near 11 to 13 profiles.

A sensitivity analysis varied k from 2 to 6 while evaluating binary recovery of the upper-severity-score category with the same fluconazole-based cluster-ranking rule. Median F1 was 0.92 to 0.98 for k = 3 to 5 at 5% to 15% prevalence. It declined to 0.14 to 0.37 at k = 2 and was unstable in some cells at k = 6, including F1 = 0.43 at 15% prevalence.

Multiclass ARI compared the full forced k = 4 partition with all four severity-score categories. Binary ARI compared the upper-severity-score category with all other categories combined. Median multiclass ARI ranged from 0.41 to 0.87 across prevalence and noise conditions and methods, with the highest values (0.85-0.87) confined to hierarchical clustering at 1-2% prevalence, coinciding with the anomalous hierarchical F1 behavior at those conditions noted above; median binary ARI was 0.026 to 0.029 at 1% prevalence and 0.93 to 0.98 at 2% prevalence and above.

## 4. Discussion

The reconstructed data rarely supported more than one natural cluster, although a forced four-group partition consistently isolated the upper-caspofungin MIC category. The gap statistic selected k = 1 in 96% to 100% of the 800 nested reconstructions. All three clustering methods recovered the prespecified category when caspofungin was included, but performance declined substantially when it was removed. Recovery was therefore driven mainly by the prominent caspofungin feature rather than by a reproducible pattern shared across all five drugs. These findings describe recovery of a caspofungin-defined category but do not establish four naturally occurring multivariate MIC phenotypes. The component counts selected by BIC also differed between the selected and extreme near-perfect-alignment reconstructions, showing sensitivity to the assumed dependence structure.

Sampling from the full published marginal distributions allowed every reported MIC category to contribute while accommodating unequal sample sizes. Reconstructing 172-profile datasets from the larger marginals, however, discarded some information and introduced sampling uncertainty. In the evaluated moderate-dependence condition, variation from pairing drug values across profiles was substantially greater than variation from resampling the marginal values. Without observed joint measurements, plausible pairings can preserve the same marginals while producing different multivariate geometry.

At low prevalence, the methods usually captured the few target profiles, so recall remained near 1.0. They also assigned many nontarget profiles to the predicted upper category, reducing precision, the F1 score, and binary ARI. Multiclass ARI summarized agreement across all four severity-score categories and was less sensitive to errors involving a very small target category. Performance improved sharply once approximately six to seven target profiles were present for GMM and k-means, with a more gradual transition for hierarchical clustering. In one low-prevalence condition, hierarchical-clustering performance increased at the highest noise level, but the accompanying bimodality does not support a general benefit of measurement noise.

The analyses identify practical conditions for recovering rare high-MIC profiles from reconstructed susceptibility data. The caspofungin-defined category was consistently recovered when caspofungin was included, while recovery of the composite severity category improved with the realized target count. This recovery should not, however, be interpreted as evidence of stable population structure.

### 4.1. Limitations

Observed joint MIC profiles for the five drugs were unavailable; each five-drug row was a synthetic pairing of values from published marginal distributions. Cross-drug dependence was imposed through a Gaussian copula at prespecified correlation levels and was not estimated from observed joint data. Therefore, the analysis cannot establish whether elevated MIC values for different drugs occur jointly in individual isolates.

The five drugs had unequal published sample sizes (172 to 581). Reducing each marginal to $n = 172$ discarded information from the four larger marginals and introduced sampling uncertainty, which was evaluated through nested resampling. Open-ended MIC bins were encoded at their boundary concentration and compared with a one-dilution-higher alternative across 400 reconstructions. The Thompson et al. sample was clinically selected rather than randomly sampled from the *Coccidioides* population, so its marginal frequencies should not be interpreted as population prevalence. Findings remained conditional on the evaluated reconstruction procedure, dependence scenarios, encoding alternatives, and Monte Carlo replication counts.

Caspofungin defined the upper-caspofungin MIC target and was included as a clustering feature; the ablation analysis showed substantially lower recovery without caspofungin. The gap statistic selected $k = 1$ in 96% to 100% of the 800 nested reconstructions, so results at $k = 4$ describe a prespecified partition. The severity-score categories were deterministic quantile bins of a researcher-defined composite score. Agreement among clustering methods does not establish biological meaning because the methods may reflect the same reconstruction assumptions or imposed geometry.

Validated *Coccidioides*-specific clinical breakpoints were unavailable for the five drugs. Terms such as upper-MIC tail and higher or lower MIC are therefore descriptive. Biological and clinical interpretation requires observed joint isolate-level data.

## 5. Conclusion

This study establishes practical conditions under which rare high-MIC profiles can be recovered from reconstructed susceptibility data. A forced four-group partition consistently recovered the caspofungin-defined threshold category when caspofungin was included, and the rare-category experiment showed that recovery improved once enough target profiles were realized. At the same time, the gap statistic selected k = 1 in 96% to 100% of nested reconstructions, demonstrating that recovery of a predefined category should not be interpreted as evidence of natural population structure. Analyses based only on marginal MIC distributions should therefore report sensitivity to the assumed cross-drug dependence and to the realized number of observations in rare target categories.

**Acknowledgements:** Summer Undergraduate Research Fellowship 2026 Opportunity provided by John Carroll University, Cleveland, OH.
**Funding:** This research received no specific grant from any funding agency in the public, commercial, or not-for-profit sectors.
**Conflicts of interest:** The authors declare no conflicts of interest.
**Author contributions:** M.W. performed the data reconstruction, simulation analyses, and drafted the manuscript. P.B. conceived and supervised the study, contributed to the analytical design, and revised the manuscript. Both authors read and approved the final manuscript.
**Data availability:** The marginal MIC frequency data underlying this study are published in Thompson et al. (reference 8, Table 2). The synthetic reconstructed datasets and simulation outputs generated for this study are available from the corresponding author upon reasonable request.
**Code availability:** The code used for this study is available from the corresponding author upon reasonable request.
**Ethics statement:** This study analyzed only synthetic MIC profiles reconstructed from previously published, de-identified aggregate frequency data. No human participants, animal subjects, or identifiable clinical records were involved, and no ethical approval was required.